\documentclass[a4paper,12pt]{article}
\usepackage[]{graphicx}
\usepackage[english]{babel}
\usepackage{mathtools} 
\usepackage{amstext}
\usepackage{amsmath}
\usepackage{amsmath,amssymb,amsthm,upref}
\usepackage{listings}
\usepackage[latin1]{inputenc}
\renewcommand{\baselinestretch}{1}

\newcommand{\e}{ \,\text{e}[}

\newcommand{\ke}[1]{ \,\text{e}[{#1}]}
\newcommand{\alp}{|}
\newcommand{\re}{]}

\newcommand{\h}[1]{\ke{-}}

\usepackage{listings,color}
\lstdefinestyle{Mastil}{%
language=Mathematica,
morekeywords={setlength},
basicstyle=\color{black}\ttfamily,
commentstyle=\color{darkgray},
stringstyle=\color{red},
xleftmargin=0pt,
xrightmargin=0pt,
backgroundcolor=\color[rgb]{1,1,1},
escapeinside={(@}{@)},
columns=fixed,
gobble=0,
tabsize=1,
keepspaces=true
number=left
texcl=true
mathescape=true
}
\lstnewenvironment{Makod2}[1][]{\lstset{style=Mastil,#1}}{}
\lstnewenvironment{Makod}[1][]{\lstset{style=Mastil,numbers=none, numberstyle=\tiny, stepnumber=1, xleftmargin=-10pt,#1}}{}
\usepackage[matrix,frame,arrow]{xy}
\usepackage{amsmath}

\newcommand{\ket}[1]{\left\vert{#1}\right\rangle}
\newcommand{\qw}[1][-1]{\ar @{-} [0,#1]}
\newcommand{\qwx}[1][-1]{\ar @{-} [#1,0]}

\newcommand{\gate}[1]{*{\xy *+<.6em>{#1};p\save+LU;+RU **\dir{-}\restore\save+RU;+RD **\dir{-}\restore\save+RD;+LD **\dir{-}\restore\POS+LD;+LU **\dir{-}\endxy} \qw}

\newcommand{\control}{*!<0em,.025em>-=-{\bullet}}

\newcommand{\ctrl}[1]{\control \qwx[#1] \qw}

\newcommand{\targ}{*!<0em,.019em>=<.79em,.68em>{\xy {<0em,0em>*{} \ar @{ - } +<.4em,0em> \ar @{ - } -<.4em,0em> \ar @{ - } +<0em,.36em> \ar @{ - } -<0em,.36em>},<0em,-.019em>*+<.8em>\frm{o}\endxy} \qw}

\newcommand{\multigate}[2]{*+<1em,.9em>{\hphantom{#2}} \qw \POS[0,0].[#1,0];p !C *{#2},p \save+LU;+RU **\dir{-}\restore\save+RU;+RD **\dir{-}\restore\save+RD;+LD **\dir{-}\restore\save+LD;+LU **\dir{-}\restore}
\newcommand{\ghost}[1]{*+<1em,.9em>{\hphantom{#1}} \qw}

\newcommand{\lstick}[1]{*!R!<.5em,0em>=<0em>{#1}}

\newcommand{\Qcircuit}[1][0em]{\xymatrix @*[o] @*=<#1>}

\author{Peter Nyman\\  
 International Center for Mathematical Modeling in Physics,\\ Engineering and Cognitive science, MSI ,Växjö University,\\ S-35195, Sweden
}
\title{Simulations of quantum error correction in classical computers}

\title{Simulation of Quantum Error Correcting Code}
\begin{document}
\maketitle

\begin{abstract}
This study considers implementations of error correction in a simulation language on a classical computer. Error correction will be necessarily in quantum computing and quantum information.  We will give some examples of the implementations of some error correction codes.
These implementations will be made in a more general quantum simulation language on a classical computer in the language \textit{Mathematica}.  The intention of this research is to develop a programming language that is able to make simulations of all quantum algorithms and error corrections in the same framework. The program code implemented on a classical computer will provide a connection between the mathematical formulation of quantum mechanics and computational methods. This gives us a clear uncomplicated language for the implementations of algorithms.
        
\end{abstract}
\section{Introduction}
The mathematical model of quantum computers is an idealization of a physical quantum computer.  In a physical quantum computer decoherence with the environment causes errors. The use of error correction provides a possibility to reduce the effect of errors. A number of mathematical models in error correction in the form of an error-correcting code have been developed. We will implement some of them in the simulation framework developed by us. This framework in Mathematica is a computer language for the simulation of quantum computers in classical computers. Within this framework we will transform the mathematical model of quantum mechanics into a computational code. Thus it will be a straightforward matter to implement quantum algorithms and error correcting codes in this language.  More specifically this means that it will represent the Dirac notation and theory connected to this notation in a natural manner. We build a state in superposition of the computational basis $\ket{0}$ and $\ket{1}$ act on this state with quantum gates. This state will be an increased $n$-qubit state with the use of the tensor product. Thus the $n$-qubit state will be a superposition of the computational basis $\ket{0}^{\otimes n}$ to $\ket{1}^{\otimes n}$. We will act with one qubit gate or two-qubit controlled-NOT  gates on this state.  Together this will give us a sufficient device for simulating quantum computers.

\section{The Simulation Framework}\label{Basicstate}
Let us introduce the part of the program that will be the framework for the simulation of error correction. This framework must naturally be part of the quantum algorithm that demands protection from error correction. In a previous study several well-known quantum algorithms have been implemented in this framework by the author (see \cite{Nyman:2007,Nyman:2007b,Nyman:2005}). We will point out that there is a symbolic similarity between our framework and the mathematical foundation of quantum computing. For this reason we will represent the code by a simple modification of Dirac's notation.
A quantum state in $n$ dimensions can be represented by a linear
combination of $n$ numbers of basis vectors as $\{\ke{0},\ke{1},\ldots\ke{n}\}=\{\ke{0}^{\otimes n},\ke{0}^{\otimes n-1}\ \otimes\ke{1},\ldots,\ke{n}\}$. In the two-dimensional case a quantum state $|\phi\rangle$ is represented as a superposition of two basis vectors, say $|0\rangle$ and $|1\rangle$, known as computational basis
(computational basis, see \cite{Hirvensalo:2001,Chuang:2000}).
In this basis a
quantum state $|\phi\rangle$ is represented as
\begin{equation}\label{phi}|\phi\rangle=\alpha|0\rangle+\beta|1\rangle,
\end{equation},
where $\alpha$ and $\beta$ are complex numbers such as $|\alpha|^2+|\beta|^2=1$.
We will introduce some new symbols for the states of the computational basis as follows:
$\text{e}[0]= |0\rangle$ and $\text{e}[1]= |1\rangle.$ This is the
foundation for the structure of the program code. For more than
one qubit we will use the computational basis states
$\text{e}[x_1,\ldots,x_n]= |x_1\ldots x_n\rangle$, where
$x_j\in\{0,1\}$ or the more compact notation $\text{e}[y]=
|y\rangle$, where $y=x_n 2^0+\dots+ x_1 2^{n-1} $. We will write
the state $\phi$ as $ \text{e}[\phi]=\alpha \text{e}[0]+\beta
\text{e}[1]$, in analogy to (\ref{phi}). The operator $A$
acts on the state $\phi$ and is usually written as $A|\phi\rangle$ in
the quantum mechanical literature. To match these symbols we will
use the computational symbols $A\alp \text{e}[\phi]$ for this
operation. One might regard $\ket{x_1\ldots\ket{y_1\ldots y_m}\ldots x_n}$ as $\ket{x_1\ldots y_1\ldots y_m \ldots x_n}$ in order to simplify the program code. This will be a computational problem, since Mathematica will distinguish between $\ke{x_1,\ldots,\ke{y_1,\ldots, y_m},\ldots ,x_n}$ and $\ke{x_1,\ldots ,y_1,\ldots, y_m ,\ldots, x_n}$.
The computer must regard these expressions as equal even if the notations are not identical with each other. As an example the expression  $\text{e}[0,\text{e}[1],1]$ must be equal to $\text{e}[0,1,1]$ in the code. We can bring in the command $\text{e}[0,\text{e}[1],1]:=\text{e}[0,1,1]$ or the more general $ \text{e}[a\_\_,\text{e}[b\_\_],c\_\_]:=\text{e}[a,b,c]$  to solve this problem. Moreover, the program code must be able to handle the linearity of
the tensor product. Let $\text{e}[\,.\,]$ be vectors and
$\alpha$ a complex number. We define the tensor product as
\begin{eqnarray}\label{tensor1}
\alpha(\text{e}[v]\otimes \text{e}[w])=(\alpha\text{e}[v])\otimes\text{e}[w]=\text{e}[v]\otimes(\alpha\text{e}[w])\\
(\text{e}[v_1]+ \text{e}[v_2])\otimes\text{e}[w]=\text{e}[v_1]\otimes\text{e}[w]+ \text{e}[v_2]\otimes\text{e}[w]\\
\text{e}[v]\otimes(\text{e}[w_1]+ \text{e}[w_2])=\text{e}[v]\otimes\text{e}[w_1]+ \text{e}[v]\otimes\text{e}[w_2].
\end{eqnarray}.
Two short commands in the program code will implement this
definition of the tensor product. The command
\begin{Makod2}
	e[a___, (@$\alpha$@)_. e[x__], b___] := (@$\alpha$@) e[a, x, b]
\end{Makod2}
\renewcommand{\baselinestretch}{1}
will transform $\text{e}[a]\otimes\alpha\text{e}[x]\otimes\text{e}[c]$  into
$\alpha\text{e}[a\otimes x\otimes b]=\alpha\text{e}[a,x,b]$. This command is the computational dual to the tensor expression in Dirac's notation  $ |a\rangle\otimes\alpha |x\rangle\otimes|b\rangle= \alpha |a\, x\, b\rangle$. The other command
\begin{Makod2}
	e[a___, (@$\xi$@)_. ((@$\alpha$@)_. e[x__] + (@$\beta$@)_. e[y__]), b___]:=
	(@$\xi\alpha$@)e[a, x, b]+ (@$\xi\beta$@)e[a, y, b]
\end{Makod2}
will transform $\text{e}[a]\otimes\xi(\alpha\e x\re+\beta\e
y\re)\otimes\text{e}[b]$ to $\xi\alpha\text{e}[a,x,b]+\xi\beta\text{e}[a,y,b]$. Let $U$ be an arbitrary unitary one-qubit quantum
gate. Then $U$ will transform a one-qubit state $\e\phi\re$, which
is represented in the computational basis states as $\e\phi\re=a\e
0\re+ b\e 1\re $, into the state $U\alp   \e\phi\re\rightarrow\ a(c_1\e
0\re+ c_2\e 1\re )+ b(c_3\e 0\re+ c_4\e 1\re )$, where $a,b,c_i$ are complex numbers. We add the \textit{Mathematica} gate $U$ to the
program code as follows: $U|\e0\re\rightarrow c_1\e 0\re+ c_2\e
1\re$ and   $U|\e1\re\rightarrow c_3\e 0\re+ c_4\e 1\re $. For
example, the Hadamard gate $H$ will be added in {Mathematica}
as the command $H{:=}\{\e0\re\rightarrow 1/\sqrt{2}(\e 0\re+ \e
1\re),\e1\re\rightarrow 1/\sqrt{2}(\e 0\re- \e 1\re)\}$. We will
define a one-qubit gate $O_i$ as an operator which acts on the qubit
in position $i$ and leaves the other qubits unchanged. The program
code must be able to operate with a gate on an arbitrary qubit.
Consequently, we will define an operator $O_i$ in the
{Mathematica} code. Defined the operator $O_i$ as $O_i=I^{\otimes i-1}\otimes U\otimes I^{\otimes n-i}$, which acts on $n$-qubits where
$I$ is the one-qubit unit operator and $U$ is an arbitrary one-qubit operator.
Then operator $O_i$ is a function of $O_i\alp \e v\re\rightarrow \e\psi\re$.
Similarly, we will define $O_{i,j}$ as an operator  which operates
as the two-qubit operator on the qubits in positions $i,j$ and
leaves the other qubits unchanged. Now we have the tools to build the quantum circuit for quantum algorithms and error correction.

\subsection{Quantum Error Correcting}\label{Quantum Error Correcting}
Quantum computers need some connection to make them controllable and will therefore never be completely isolated from their surroundings.  The quantum computer's surroundings will influence the quantum computer and cause errors. The effects of errors caused by inaccuracy and decoherence need to be reduced to a minimum. Error correction will help to reduce these effects. Define the 3-qubit "logical qubits" denoted  $\ket{1_L}$ (logical-one) and $\ket{0_L}$ (logical-zero), as it is commonly defined in the literature 
\begin{equation}\label{Firstlogicalqubits}
\ket{0_L}=\ket{000},\quad
\ket{1_L}=\ket{111} .
\end{equation}
Consider some error that will cause a flip on one of the qubits in the logical qubits (e.g. $\ket{000}$ change to $\ket{010}$).  This qubit flip will be represented by the Pauli operator $X_i$, where $i$ denotes the position of the flipped qubit. Thus we will define $X_1=X\otimes I\otimes I$, $X_2=I\otimes X\otimes I$ and $X_2=I\otimes I\otimes X$.  In the same manner a phase flip $\ket{1}\mapsto -\ket{1}$ and the combination of phase flip and qubit flip will be represented by the Paul operator $Z_i$ and $X_iZ_i=-iY_i$, respectively. Moreover, ignore the global phase\footnote{Two states which are equal up to the global phase will be equal in the observer's eyes (see: \cite[p.93]{Chuang:2000})}
and denote a combination of a qubit and phase flip as $Y_i$. A linear combination of the Pauli matrices and the identity matrix will then represent an error operator 
\begin{equation}\label{Error}
	E_i=c_1I_i +c_2X_i +c_3Z_i +c_4Y_i.
\end{equation}
 
 The use of majority voting and a preparation of the state in logical qubits \eqref{Firstlogicalqubits} will protect the state against errors that flip a single qubit. This error code will not protect against a phase flip. The Shor code \cite{Shor:1995} will overcome this problem. Therefore prepare the state in the logical qubits 
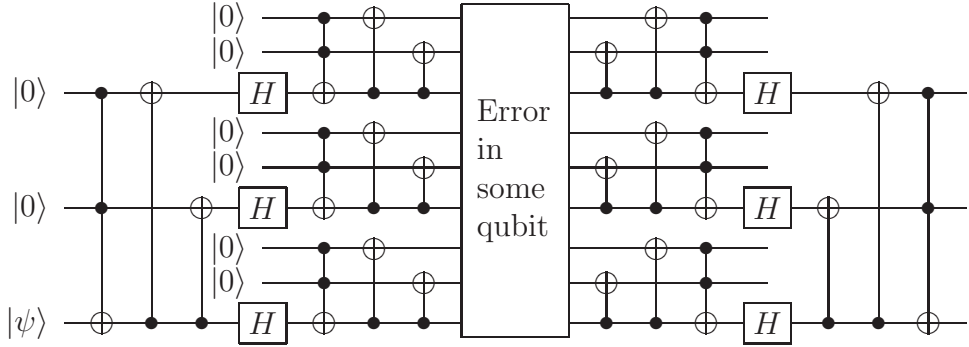
\begin{figure}
	\centering
\[\Qcircuit @C=0.8em @R=0.2em {%
	&		&		&		&	\lstick{\ket{0}}	&	\ctrl{1}	&	\targ	&	\qw	&	\multigate{8}{\begin{minipage}[l]{1.cm}{Error\\in\\some\\qubit}\end{minipage}}	&	\qw	&	\targ	&	\ctrl{1}	&	\qw	&		&		&		&		\\
	&		&		&		&	\lstick{\ket{0}}	&	\ctrl{1}	&	\qw	&	\targ	&	\ghost{{error}}	&	\targ	&	\qw	&	\ctrl{1}	&	\qw	&		&		&		&		\\
\lstick{\ket{0}}	&	\ctrl{6}	&	\targ	&	\qw	&	\gate{H}	&	\targ	&	\ctrl{-2}	&	\ctrl{-1}	&	\ghost{{error}}	&	\ctrl{-1}	&	\ctrl{-2}	&	\targ	&	\gate{H}	&	\qw	&	\targ	&	\ctrl{6}	&	\qw	\\
	&		&		&		&	\lstick{\ket{0}}	&	\ctrl{1}	&	\targ	&	\qw	&	\ghost{{error}}	&	\qw	&	\targ	&	\ctrl{1}	&	\qw	&		&		&		&		\\
	&		&		&		&	\lstick{\ket{0}}	&	\ctrl{1}	&	\qw	&	\targ	&	\ghost{{error}}	&	\targ	&	\qw	&	\ctrl{1}	&	\qw	&		&		&		&		\\
\lstick{\ket{0}}	&	\ctrl{3}	&	\qw	&	\targ	&	\gate{H}	&	\targ	&	\ctrl{-2}	&	\ctrl{-1}	&	\ghost{{error}}	&	\ctrl{-1}	&	\ctrl{-2}	&	\targ	&	\gate{H}	&	\targ	&	\qw	&	\ctrl{3}	&	\qw	\\
	&		&		&		&	\lstick{\ket{0}}	&	\ctrl{1}	&	\targ	&	\qw	&	\ghost{{error}}	&	\qw	&	\targ	&	\ctrl{1}	&	\qw	&		&		&		&		\\
	&		&		&		&	\lstick{\ket{0}}	&	\ctrl{1}	&	\qw	&	\targ	&	\ghost{{error}}	&	\targ	&	\qw	&	\ctrl{1}	&	\qw	&		&		&		&		\\
\lstick{\ket{\psi}}	&	\targ	&	\ctrl{-6}	&	\ctrl{-3}	&	\gate{H}	&	\targ	&	\ctrl{-2}	&	\ctrl{-1}	&	\ghost{{error}}	&	\ctrl{-1}	&	\ctrl{-2}	&	\targ	&	\gate{H}	&	\ctrl{-3}	&	\ctrl{-6}	&	\targ	&	\qw	
}\]																																	
	\caption{Circuit for encoding and decoding Shor code}
	\label{Shor code}
\end{figure}
 \begin{equation}\label{shorlogicalqubits}
\ket{0_L}=\left(\frac{\ket{000}+\ket{111}}{\sqrt{2}}\right)^{\otimes 3},  \quad
\ket{1_L}=\left(\frac{\ket{000}-\ket{111}}{\sqrt{2}}\right)^{\otimes 3}.
\end{equation}
The encoding of the logical states in Shor's code is presented in the left part of the circuit in figure \ref{Shor code}. Some arbitrary error will be simulated by the error operator \ref{Error} in the middle of the circuit. This operator simulates a one-qubit error which can be a phase or a flip error or some combination of these two. The decoding is the inverse to encoding, as will be seen in the circuit.

\section{The simulation}
The representations of error correcting code (Shor's code) simulation in Mathematica will follow in this section. First define the quantum computer properties by the code in the listing
\begin{Makod}[caption=Definition of register and quantum gates in Mathematica]
	e[a___,(@$\alpha$@)_.e[x__],b___]:=(@$\alpha$@)e[a,x,b]
	e[a___,(@$\xi$@)_.((@$\alpha$@)_.e[x__]+(@$\beta$@)_.e[y__]),b___]:=
			(@$\xi\alpha$@)e[a,x,b]+(@$\xi\beta$@)e[a,y,b]
	e[a___,e[],b___]:=e[a,b]
	O(@$_{\_i_{\_}}|$@)v_:=Chop[Expand[v/.
			(e[x__](@$:\to$@)ReplacePart[e[x],e[{x}[(@$\!\!$@)[i](@$\!\!$@)]]/.O,i])]]
	O(@$_{\_i_{\_},j_{\_}}|$@)v_:=Chop[Expand[v/.O[i,j]]
	O(@$_{\_i_{\_},j_{\_},k_{\_}}|$@)v_:=Chop[Expand[v/.O[i,j,k]]]
	(@\bfseries{CN}@)[i_,j_]:={e[x__](@$:\to$@)e[e[x][[1;;i-1]],
			e[Sequence@@Mod[e[x][[i]]+e[[j]],2]],
			    e[x][[i+1;;-1]]]}
	(@\bfseries{T}@)[i,j,k] := {e[x__](@$:\to$@)e[e[x][[1;;k-1]],
	    e[Sequence@@Mod[e[x][[k]]+(e[x][[i]]*e[[j]]),2]],
	        e[x][[k+1;;-1]]]}
	(@\bfseries{H}@) := {e[0](@$:\to 1/\sqrt{2}$@)(e[0]+e[1]),e[1](@$:\to 1/\sqrt{2}$@)(e[0]-e[1])}
	(@\bfseries{X}@) := {e[0](@$:\to$@)e[1],e[1](@$:\to $@)e[0]}
	(@\bfseries{Y}@) := {e[0](@$:\to$@)-i e[0],e[1](@$:\to $@)i e[1]}
	(@\bfseries{Z}@) := {e[0](@$:\to$@)e[0],e[1](@$:\to $@)-e[1]}
	(@\bfseries{Id}@) := {}
\end{Makod}
This part of the code has defined the register and the quantum gates, and the simulation is ready for the quantum circuit. The next part, which expresses the quantum circuit, is divided into encoding, simulating errors, decoding and measuring. The Enlarge function in the listing 2 will be a simulation of the extension of an arbitrary computational one-qubit state to a nine-qubit state. The encoding will affect the Shor code, which is easy to compare with the encoding in a quantum circuit \ref{Shor code}. Hence read the Mathematica code from inward out. The first quantum gate to implement is the $\textbf{CN}_{4,1}$ gate, the next one the $\textbf{CN}_{7,1}$ gate, and so forth. The simulation of noise is implemented by the Error function. After applying some noise we can measure the state and return to the initial state.    
\begin{Makod}[caption=Encode and decode] 
Enlarge[(@$\psi$@)_,i_]:=(@$\psi$@)/.e[x ](@$:\to$@)e[x,Sequence@@Table[0,{i}]]
Encoding[(@$\psi$@)_]:=(@$\textbf{CN}_{9,7}$@)|((@$\textbf{CN}_{8,7}$@)|((@$\textbf{CN}_{6,4}$@)|((@$\textbf{CN}_{5,4}$@)|((@$\textbf{CN}_{3,1}$@)|((@$\textbf{CN}_{2,1}$@)|((@$\textbf{H}_{7}$@)|((@$\textbf{H}_{4}$@)|((@$\textbf{H}_{1}$@)|
        ((@$\textbf{CN}_{7,1}$@)|((@$\textbf{CN}_{4,1}$@)|(@$\psi$@)))))))))))) 
Error[(@$\psi$@)_,i]:=(@$a_i$@)((@$Id_i$@)|(@$\psi$@))+(@$b_i$@)((@$X_i$@)|(@$\psi$@))+(@$d_i$@)((@$Z_i$@)|(@$\psi$@))+(@$d_i$@)((@$Y_i$@)|(@$\psi$@))
Decoding[(@$\psi$@)_]:=(@$\textbf{T}_{7,4,1}$@)|((@$\textbf{CN}_{7,1}$@)|((@$\textbf{CN}_{4,1}$@)||((@$\textbf{H}_{7}$@)|((@$\textbf{H}_{4}$@)|((@$\textbf{H}_{1}$@)|((@$\textbf{T}_{9,8,7}$@)|
         ((@$\textbf{CN}_{9,7}$@)|((@$\textbf{CN}_{8,7}$@)|((@$\textbf{T}_{6,5,4}$@)|((@$\textbf{CN}_{6,4}$@)|((@$\textbf{CN}_{5,4}$@)|((@$\textbf{T}_{3,2,1}$@)|((@$\textbf{CN}_{3,1}$@)|((@$\textbf{CN}_{2,1}$@)|(@$\psi$@))
         ))))))))))))) 
Measure[(@$\psi$@)_]:=FullSimplify[(@$\psi$@)/.e[y_,x__](@$\to$@)e[y]]
	\end{Makod}
Let us encode the arbitrary computational initial state $\psi_0$ and encode by the Shor code.
The simulation of a $i:th$ qubit error on the $\psi_1$ state is implemented by the Enlarge function [$\psi_1$,i] (specific to this example, an error in qubit $8$).
Finally, decode and measure the state.
	\begin{Makod}[caption=Algorithm]
(@$\psi_0$@) = (@$\alpha$@)e[0]+(@$\beta$@)e[1];
(@$\psi_1$@) = Enlarge[(@$\psi_0$@),9];
(@$\psi_2$@) = Encoding[(@$\psi_1$@)];
(@$\psi_3$@) = Error[(@$\psi_2$@),8];
(@$\psi_4$@) = Decoding[(@$\psi_3$@)];
(@$\psi_5$@) = Measure[(@$\psi_4$@)]	
\end{Makod}
The output will be $(\alpha e[0]+ \beta e[1])\left(a_8+b_8+c_8-i d_8\right)$, where the global phase $\left|a_8+b_8+c_8-i d_8\right|^2=1$if the error operator is a unitary operator.
In fact, the global phase can be ignored and the state $(\alpha e[0]+ \beta e[1])\left(a_8+b_8+c_8-i d_8\right)$ and $(\alpha e[0]+ \beta e[1])\left(a_8+b_8+c_8-i d_8\right)$ will be considered as equal in an observational point.

\end{document}